\documentclass[AMA,Times1COL]{WileyNJDv5} 

\makeatletter
\def\oddhead@titlepage@info{}
\makeatother

\articletype{Research Article}%

\journal{Preprint}
\volume{xx}
\copyyear{2025}
\startpage{1}

\begin{document}

\title{Impacts of Blade Camber on Cross-Flow Turbine Performance and Loading}

\author[1]{Ari Athair}

\author[2]{Caelan Consing}

\author[2]{Jennifer A. Franck}

\author[1]{Owen Williams}

\authormark{ATHAIR \textsc{et al.}}
\titlemark{Impacts of Blade Camber on Cross-Flow Turbine Performance and Loading}

\address[1]{\orgdiv{William E. Boeing Department of Aeronautics and Astronautics}, \orgname{University of Washington}, \orgaddress{\city{Seattle},\state{WA}, \country{U.S.}}}

\address[2]{\orgdiv{Department of Mechanical Engineering}, \orgname{University of Wisconsin-Madison}, \orgaddress{\city{Madison},\state{WI}, \country{U.S.}}}

\corres{Corresponding author Ari Athair, \email{aristone@uw.edu}}

\fundingInfo{U.S. Department of Defense Naval Facilities Engineering Systems Command (N0002421D6400/N0002421F8712) and Department of Energy TEAMER program (EE0008895).}

\abstract[Abstract]{Cross-flow turbines show promise for renewable energy generation from wind and tidal sources. The rotating reference frame of cross-flow turbine blades results in virtual camber and incidence due to streamline curvature, altering the lift, drag and pitching moment of the blades. Adding geometric camber is therefore likely to alter performance and loading, however there is little consensus regarding the direction of camber that might be most favorable. This study compares 2\% concave-in and concave-out cambered blades (NACA 2418) with symmetrical NACA 0018 foils for a turbine with a 0.49 chord-to-radius ratio. Experimental performance measurements are compared across a range of tip-speeds, and particle image velocimetry is used to explore the in-rotor flow evolution through the cycle. Concave-out blades, which enhance virtual camber and lift in the power stroke are found to exhibit sub-optimal performance. In contrast, concave-in cambered blades slightly improved symmetrical blade performance by enhancing downstream flow reattachment, more than compensating for reduced peak power generation. The difference between each cambered foil is seen to grow with increasing tip-speed ratio. Moreover, these concave-in blades reduce peak loading by 13\%, which may prove critical in future designs, especially at high tip-speed ratios. Exploration of the near-blade flow fields suggest that the influence of geometric camber is non-linear, and that use of a simplistic summation of both geometric and virtual camber to account for camber effects may be overly simplistic. Despite this, corresponding validated simulations suggest that a small but positive total camber (geometric plus virtual) is optimal for this turbine. }


\begin{figure}
    \centering
   \includegraphics[width=9.5cm]{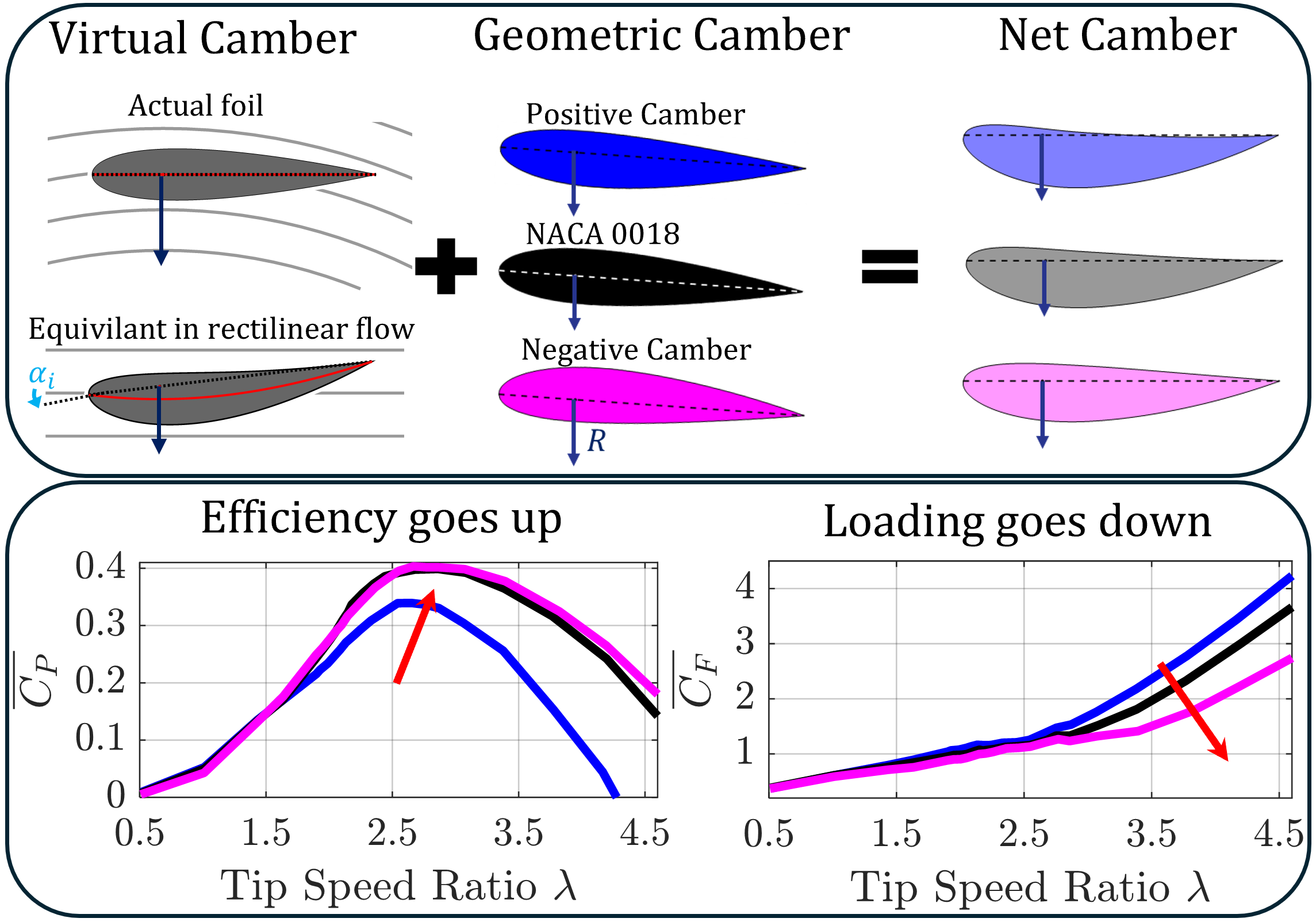}
    \caption*{Graphical Abstract}
\end{figure}%



\maketitle

\section{Introduction}

Cross-flow turbines, often referred to as vertical axis wind turbines or VAWTs, represent a promising technology for renewable energy generation from both wind and water currents. Research has shown many potential benefits, including higher power per unit area than axial flow turbines, direction-agnostic operation, reduced environmental impacts, confinement-exploiting capabilities, and lower maintenance costs \cite{Dabiri2011,Zhao2022,hunt2025HighBlockageArray}. Traditionally, symmetrical foils have been the favored choice for cross-flow turbines, as this geometry was thought to effectively accommodate the large excursions in positive and negative angle of attack ($\alpha$) experienced in such turbines \cite{Healy1978} and there has been little information to inform alternatives. Geometric camber offers a path to increased blade lift, however its interaction with virtual blade curvature, or "virtual camber" induced by the blade's rotational path is unclear, leading to uncertainty regarding the direction or magnitude of camber that might be beneficial. Similarly, the interaction of camber with, dynamic influences such as stall and flow recovery, may be significant.  This paper explores the potential benefits of geometrically cambered foils for cross-flow turbines and VAWTs, including an examination of the camber direction that leads to performance enhancement and a theoretical basis for foil optimization of generalized turbines. 

\begin{figure}
    \centering
    \includegraphics[width=0.75\textwidth]{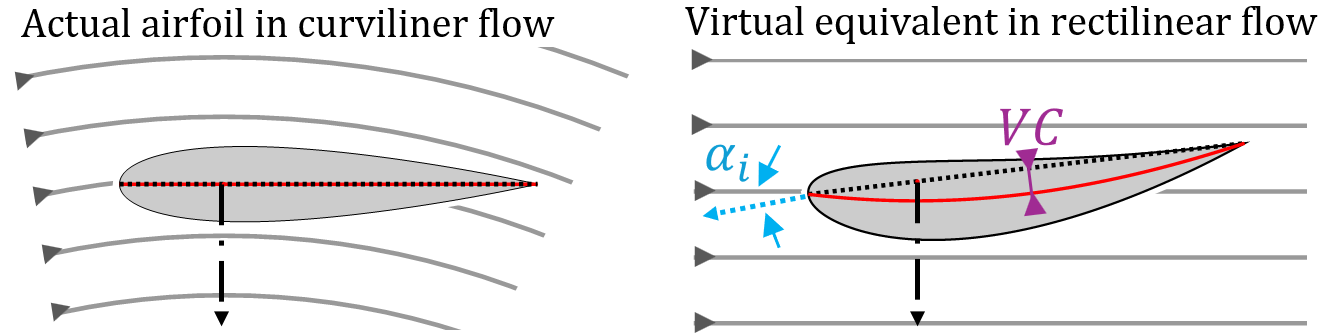}
    \caption{Schematic of flow-curvature in the blade reference frame of cross-flow turbines which result in virtual camber (VC) and nose-in pitching behavior, virtual incidence $\alpha_i$. }
    \label{fig:flowcurve}
\end{figure}

The selection of cross-flow turbine blade camber and pitch is complicated by the blade rotation. As cross-flow turbine blades move on a curved trajectory about their axis, a chord-wise variation in $\alpha$ is induced, which can be modeled as an equivalent "virtually cambered" foil in a rectilinear flow (see figure \ref{fig:flowcurve}) \cite{Migliore1980}. The angle of attack of this virtual foil is also altered by an amount known as the "virtual incidence" ($\alpha_i$). The virtual camber causes the aerodynamic behavior of turbine blades to resemble that of cambered foils at a shifted incidence, resulting in greater lift generation than expected from a symmetrical foil in a rectilinear flow.
The apparent flow curvature experienced by a blade rotating about an axis is dependent on its chord-to-radius ratio, $c/R$, mounting point, and, to a lesser extent, its pitch, with $c/R$ being by far the most significant \cite{Migliore1980}. Given the above definitions, the virtual camber of cross-flow turbine blades is oriented such that the virtual equivalent blade has an outward concavity (opposite the arc of rotation) throughout the cycle. We define this to be positive or ``concave out'' camber, as it increases lift in the upstream portion of the rotation, and the opposite direction to be negative or ``concave in'' camber.

\begin{figure}
    \centering
    \includegraphics[width=.95\textwidth]{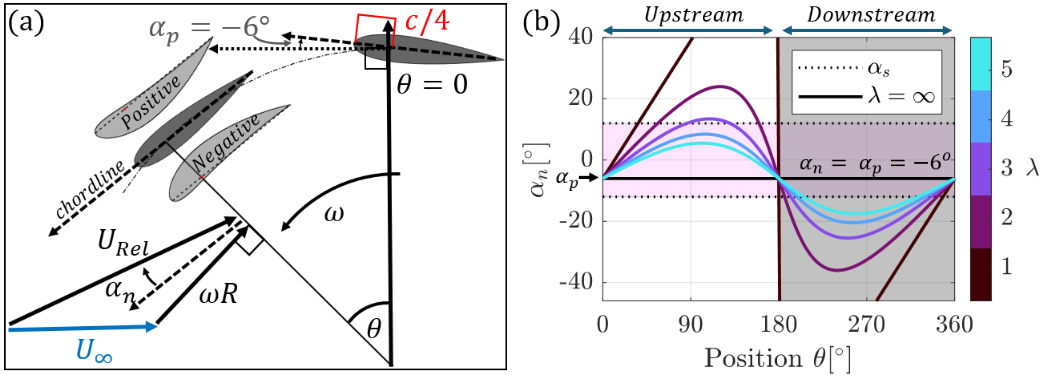}
    \caption{(a) Schematic of turbine coordinate system and key parameters. (b) Nominal angle of attack experienced by the blade throughout the cycle as a function of $\lambda$. The solid black line represents the infinite $\lambda$ condition where $\alpha_n = \alpha_p$, the gray shaded rectangle indicates the downstream region, and shaded pink denotes conditions bounded by the static stall angle $\alpha_{s}=\pm 12^\circ$ for the blades and Reynolds numbers used in this work for reference \cite{timmer2008AoA,bianchini2016experimentalAOA}.}
    \label{fig:Coordinate}
\end{figure}

During turbine operation, where rectilinear inflow is superimposed upon the curved rotating frame, the blade virtual camber and incidence also become dependent on the angular position of the blade, and the tip-speed ratio, $\lambda = \omega R/U_\infty$, where $\omega$ is the angular velocity of the turbine, $R$ is the turbine radius and $U_\infty$ is the free stream velocity far upstream of the turbine. In the current study, the blade angular position, $\theta$, is defined to be zero when the blade is pointed directly upstream (see figure \ref{fig:Coordinate}a). The azimuthal dependence of virtual geometry and incidence is a confounding factor that prevents the direct evaluation of virtual turbine geometry in rectilinear flows. As a result, the virtual camber at $\lambda = \infty$   (a purely rotational flow) is often utilized as a singular metric of this rotational influence when comparing turbine geometries.

 To predict the role of camber in turbine performance, in the context of this azimuthally varying flow environment, theoretical analyses offer a useful starting point for evaluating how both virtual and geometric camber affect blade performance. Turbine blades have a positive nominal angle of attack ($\alpha_n$) in the upstream portion of their rotation $(0< \theta< 180)$, shown in figure \ref{fig:Coordinate}b, during which the blades produce peak torque and power due to lift. Positive camber (virtual or geometric) is consequently expected to enhance lift and, therefore, torque and overall efficiency. In contrast, negative camber would be expected to reduce the peak blade lift and so might be expected to be detrimental to overall turbine efficiency. It might be considered advantageous to physically camber the foil to increase lift in the power-producing region, as other optimization studies have focused on, using a positively cambered blade to accentuate the virtual camber due to the blade's rotation \cite{Hand2021}. 
 
 While this is a promising hypothesis, there are many reasons for caution. Positive camber is also known to increase drag and pitching moment, both of which typically reduce turbine efficiency. However, these concerns based on steady-state aerodynamics may not apply to the unsteady, dynamic stall and flow recovery that dominate cross-flow turbines. Studies on pitching foils have indicated that increasing positive camber has been found to delay dynamic stall, shift stall vortex formation towards the trailing edge during upward pitch (synonymous with the upstream), and improve flow reattachment \cite{mccroskey1981phenomenon,Ouro2018,Shum2024}, each of which is expected to improve turbine performance. However, the combined effects of virtual camber and forces from the rotating frame of cross-flow turbines remain unexamined.

 While pitching foil studies offer insights to upstream blade behavior, the flow in the downstream portion of the rotation $(180< \theta< 360)$ remains less understood. In this region, non-linear flow recovery processes, reduced freestream momentum due to induction, and blade-wake interactions are present, all of which will be influenced by blade camber. Recent work by Snortland et al.\cite{snortland2025downstream} has highlighted the importance of this often underappreciated portion of the rotation to overall performance, and the strong influence of foil drag and pitching moment, especially at high $\lambda$. Thus, choosing a blade profile that improves upstream lift (positive camber) or, conversely, one that reduces drag and pitching moment downstream (negative camber) may be optimal, but there is no current theory that can determine which influence will win out.

As a result, a wide range of camber magnitudes and directions (inward or outward) have been suggested to be optimum in the literature with little agreement \cite{letcher2017wind}. Some numerical works suggest negatively cambered blades enhance performance by counteracting virtual camber \cite{Danao2012,Danao2016MoreCambers,abotaleb2024impact}, others support positive camber for improved self-starting \cite{Baker1983,Kirke1991} or wake recovery \cite{Zanforlin2023}, while others maintain minimal camber or symmetrical is superior \cite{Wang2018}, and so differences in optimization objective may also play a role in the varied conclusions. These differences may also be the result of the wide range of $c/R$ employed in previous studies without consideration of the coupled change in virtual camber. In addition, differences in blade preset pitch and mounting point make a direct comparison between prior work difficult, while also having secondary impacts on the virtual camber. The concept of “effective” or net camber---defined as the sum of geometric and virtual camber---has been proposed but not rigorously explored \cite{Kirke1998}. This leaves key questions unresolved about camber’s role across a broader design space and how it interacts with $c/R$ and other turbine parameters.

Optimizing turbine blades requires consideration of both geometric and virtual camber in concert, but this interaction has not been thoroughly studied. This work experimentally evaluates their combined effects on a cross-flow turbine with a moderate $c/R$ of 0.49 across a wide range of $\lambda$. An assessment of virtual camber and angle of attack is conducted using the method of Migliore and Wolfe \cite{Migliore1980}. We investigate whether virtual camber enhances or diminishes performance by comparing blades that reduce the effective camber and others that enhance it. Particle image velocimetry (PIV) is employed to explore near-blade hydrodynamic influences on performance in the upstream and downstream regions and comparisons are made with two-dimensional unsteady Reynolds-averaged Naiver-Stokes (URANS) simulations of the same turbine to assess performance sensitivity over a wider range of geometric camber. It is shown that optimum blade camber is a function of $\lambda$, and enhancement of downstream flow recovery through a negative camber (which reduces power stroke efficiency) is often preferred.

\section{Experimental Methods}
\label{sec:CamberMethods}
This work was conducted at the University of Washington Alice C. Tyler Flume, which has a test section 4.6 m length and 0.76 m width. This facility uses a pool heater to maintain a constant water temperature of 29 $^\circ$C and therefore Reynolds number. The inflow velocity was 0.885 m/s with a turbulence intensity of between 2 and 3\%, as measured five diameters upstream of the turbine position by an acoustic Doppler velocimeter (ADV), Nortek Vectrino. The ADV data provided a measure of the free stream velocity, $U_\infty$, that is used to normalize non-dimensional turbine performance. It was collected at a 16 Hz sample rate and assumed to be free of turbine induction effects. The dynamic depth of the flume was held at  57.2 cm, measured at a 0.5 Hz sample rate by an ultrasonic free surface transducer (Omega LVU30 Series), and the turbine was centered in the water column sufficiently downstream of the inlet to experience uniform flow. 

\begin{figure}[!t]
\centering
{\includegraphics[width=0.65\textwidth]{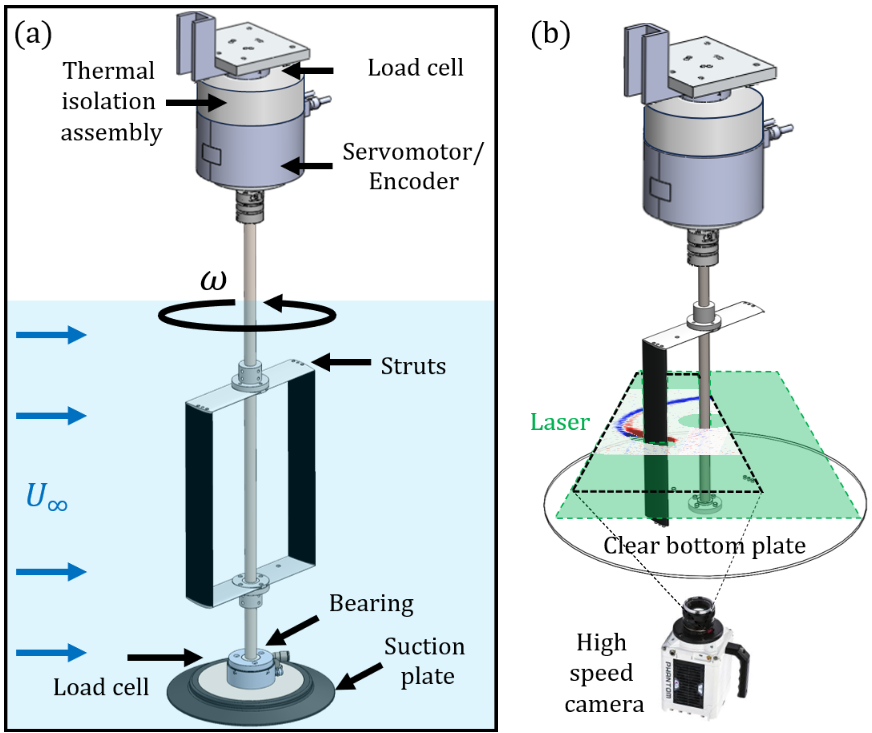}
\label{Perf_Turbine}}
\caption{(a) Experimental configuration used to collect force and performance data. (b) Cantilevered experimental setup used to capture in rotor mid-span planner PIV with representative flow field.}
\label{fig_TurbSetUp}
\end{figure}%

\subsection{Experimental Turbine Test Setup} 

The turbine used for performance and force analysis can be seen in Figure \ref{fig_TurbSetUp}a. It is a double-pinned configuration, with two six-axis ATI load cells, Mini45, mounted at the top and bottom of the central shaft that measure forces and moments at 1 kHz. A 5 Nm servo motor (Yaskawa SGMCS-05B3C41) rotated the turbine under constant speed control and recorded position and velocity throughout the cycle. This servo contains an internal encoder with resolution 2$^{18}$ counts per revolution. All turbine data was streamed through a National Instruments data acquisition system and recorded using \textit{Matlab Simulink Desktop Real-time}. For each $\lambda$, data was collected for 45 s, the necessary time to achieve convergence of the average power and remove the influence of any cycle-to-cycle fluctuations. To remove high-frequency motor and EMF noise, a low-pass Butterworth filter with a pass frequency of 10 Hz and a cutoff frequency of 100 Hz was applied to the extracted turbine data. This filter was found to have no impact on time-averaged performance \cite{SNORTLAND2025Force}.

Both one- and two-bladed turbines were tested with three different blade cambers; a  symmetrical NACA 0018, and a NACA 2418 that could be mounted in either the positive or negative camber directions. 
The mounting and dimensions of the blades remained uniform across all three configurations; with a chord length ($c$) of 4.06 cm, a span ($h$) of 23.4 cm, and preset pitch angle ($\alpha_p$) of $-$6$^\circ$, rotated leading edge outward about the quarter chord, a configuration previously found to be optimal for a symmetrical blade \cite{Hunt2024Geometry}. A 1.27 cm diameter central shaft connected the turbine assembly to the motor. The turbine radius to the quarter chord, $R$, was 8.23 cm, resulting in a chord-to-radius ratio of 0.49 and chord- and freestream-based Reynolds number, $Re_c = 44\times10^3$. Froude numbers based on the water depth, $H$, and submergence (distance to the top of the turbine) $S$, were $Fr_H = U_\infty/\sqrt{g H}= 0.37$ and $Fr_S=U_\infty/\sqrt{g S} = 0.67$, respectively.  These non-dimensional numbers were held constant across all tests to alleviate any secondary changes to performance across experiments \cite{Ross2022}.  The turbine blockage, or the ratio of the turbine projected frontal area to the flume cross-sectional area, $\beta$, was 9.3\%. The turbine solidity was $\sigma = \frac{N c}{2\pi R} = 0.16$ for the two-bladed turbine and $\sigma$ of $0.08$ for a single-bladed configuration.
The radius to the quarter chord for each blade was held constant; however, the outermost swept radius of the three blades, $R_{o}$, had slightly different values of 8.55, 8.60, and 8.66 $cm$ for the positive, symmetrical, and negatively cambered blades, respectively. These different radii were incorporated into the area calculation used to non-dimensionalize power and force coefficients, consistent with previous work \citep{Hunt2024Geometry}.

\subsection{Nondenominational Performance Metrics}
\label{sec:KeyPerfMet}
Turbine loading and torque measurements are post-processed to determine a range of key performance metrics. The turbine efficiency, or power coefficient, is defined as the ratio of extracted mechanical power,  $\tau \omega$, to the kinetic power present in the water flowing through the cross-sectional area of the turbine such that, 
\begin{equation}
                C_P(\theta) = \frac{\tau(\theta) \omega}{\frac{1}{2} \rho A U_\infty^3}  
\end{equation}
where  $\rho$ is the fluid density, $A = 2R_{o}h$ is the turbine projected area, and $\tau$ is the torque. Similarly, the horizontal force coefficient is defined as an instantaneous normalized vector sum of the lateral force, $F_{Y}$ (perpendicular to the freestream and the turbine central shaft, $y$ direction), and the thrust force, $F_{X}$ (aligned with the free stream, $x$ direction), such that, 
 \begin{equation}
                C_{F}(\theta) = \frac{\sqrt{F_{X}(\theta)^2+F_{Y}(\theta)^2}}{\frac{1}{2} \rho A U_\infty^2}. 
\end{equation}
Phase-averaged performance was calculated by bin-averaging performance data using 2$^\circ$ bins across the 45 $s$ of data recorded for each experiment. Time-averaged versions of all metrics are chosen to encompass an integer number of rotations and denoted by an overbar, e.g., $\overline{C_P}$. 

To isolate blade-level hydrodynamic efficiency and loading, the inertial effects (for the imbalanced one-bladed turbine only) and then support structure losses were subtracted from the measured phase-averaged turbine performance as obtained through supplementary experiments. Support structure losses were measured with the blades removed, as were forces and moments on an isolated center shaft under matching operating conditions. A modified superposition support subtraction developed by Strom et al. \cite{strom2018SuportStruct} and Snortland et al. \cite{SNORTLAND2025Force} was utilized to calculate the blade-level contributions to overall turbine forces and moments. Losses from the struts and central shaft are directly subtracted from total turbine measurements for thrust and torque to isolate the blade performance. The support losses of the lateral force included a physics-based weighting of bladeless experiments to accurately remove the Magnus force of the center shaft and induction within the turbine, shown to be important in this direction \cite{SNORTLAND2025Force}. While we acknowledge a higher degree of uncertainty for loading estimates, we maintain high confidence in blade-level power estimates. A detailed description of the performance subtraction method can be found in the Appendix \ref{sec:Supplimental}.

\subsection{Flow Visualization (PIV)}
A single Phantom V641 high-speed camera with 2560 x 1600 resolution and a 55 mm focal length lens captured images within a plane normal to the blades at their center span, shown in figure \ref{fig_TurbSetUp}b. The flow was illuminated by a 30 mJ per pulse Continuum Terra PIV Nd:YLF laser, producing a light sheet approximately 2 mm thick.  PIV was acquired at 30 phases with a 12$^\circ$ separation, with timing controlled by the \textit{MATLAB Simulink} turbine controller to ensure phase consistency of the acquisition. At each phase, 85 image pairs were captured. These replicates were captured for each experiment and FoV so that low-noise phase-averaged flow fields could be reconstructed and analyzed. The inter-frame time was chosen to be 350-370 $\mu$s, so maximum in-plane displacements were limited to approximately 9 pixels such that vortex cores could be sufficiently resolved. The image calibration was 10.16 pixels/mm.  Auxiliary experiments, run under matching conditions and with ambient illumination allowed the turbine central shaft to be tracked as a function of phase and small precession to be eliminated prior to processing of the PIV images. This approach follows the method developed by Snortland et al. \cite{snortland2023cycle2cycle}. 

Two mirrors positioned on the opposite side of the flume were used to reflect the laser sheet into the rotor to reduce the extent of shadowed regions and improve data yield. Fields of view (FoV) of 15.7$\times$25.1 cm$^2$ were collected by maintaining the same laser position and changing the relative turbine position to collect data in multiple regions relative to the turbine after measuring to a datum. Two FoVs were collected, one capturing the upstream blade path, and the other, the downstream. FoVs were overlapped by approximately 3 cm to create a larger,  synthesized phase-averaged domain.

PIV data were analyzed using \textit{LaVision DaVis 10.2} software employing a multi-grid, multi-pass approach with iterative image deformation. Butterworth high-pass background subtraction was applied to a 49 frames moving block of images as a prepossessing step to remove background illumination. Hand-drawn masks were created for each phase to remove non-illuminated regions, such as the central shaft, blade shadows and any remaining poorly illuminated regions. The interrogation window size of the final pass was 32$\times$32 pixels with 75\% overlap. A vector post-processing step was applied three times to remove erroneous vectors. It involved a universal outlier detection filter on a 5$\times$5 vector domain with a threshold of three standard deviations. The resulting processed data was exported and phase-averaged for further plotting and analysis in MATLAB.

\subsection{Numerical setup}

Simulations are performed to explore a wider range of camber geometries. The governing equations are the incompressible unsteady Reynolds-averaged Navier-Stokes (URANS) equations, solved in two-dimensions. The solver is implemented using the second-order finite-volume spatial discretization in OpenFOAM.  The Reynolds stress tensor is modeled with $k$-$\omega$ shear stress transport (SST) equations.

The meshing, numerical solver, and turbine set-up for the computations very closely follows previous turbine simulations validated with experimental force and PIV data \cite{dave2021simulations, dave2021comparison}. The only deviations from these prior works is that the current simulations are single-bladed turbines, and each blade has camber. Single-bladed turbines allow for more in-depth examination of the blade-level forces and their impact on power, and enable one-to-one comparison with the single-bladed turbine experiments. 

The computational turbine has $c/R=0.49$ with an outwards pitch of 6 degrees, matching the experiments but does not consider the struts or center shaft. The two-dimensional domain has a blockage of 13\%, with slip walls along the top and bottom.  The computational domain extends 300 chord lengths in the streamwise direction with the turbine placed in the center, and has incompressible inlet and outlet conditions. The Reynolds number based on chord and freestream is 45,000.

The baseline blade shape is the same as the experiments, a NACA 0018, a symmetric blade with 18\% maximum thickness. The mesh surrounding the airfoil is structured and body-fitted. All URANS models are sensitive to mesh resolution, especially around the surface of an airfoil with flow separation. Thus, camber was generated via a mesh-motion algorithm that deformed the baseline mesh to the desired camber, fully documented in prior work \cite{Bridges2022thesis}. This eliminates the need for remeshing and ensures that the near-wall resolution remains constant for an equal comparison across various cambers.

\subsection{Estimation of Virtual Camber and Incidence}

Migliore and Wolfe \cite{Migliore1980} were the first to propose an approach to determine the virtual camber of airfoils in straight-bladed Darrieus (cross-flow) wind turbines, utilizing a geometric transformation. Subsequently, a number of other researchers have developed alternative approaches. Extensions to include the effects of active pitching-induced velocity variation were developed by two groups \cite{Cardona1984, Halder2017}. Others used alternative mathematical approaches to the transformation that have been both more straightforward, such as using a virtual camber line of a circular arc \cite{Zervos1988,Mandal1994}, and more complex, utilizing exponentials and conformal transformations \cite{Akimoto2013}. Similar methods have been developed for analysis of submarines making constant radius turns and specialized wind tunnels to evaluate these curved flows for automobile cornering, illustrating the breadth of impacts of curved flows \cite{Gregory2004,keogh2015techniques}.  These approaches are commonly evaluated and compared in the limit of infinite $\lambda$, when the virtual geometry becomes constant with blade position. In this limit, it has been shown that all result in similar predictions for  $c/R<0.4$.  For larger $c/R$, the results of these methods diverge slightly \cite{Horst2016}. 

It should be noted that all virtual camber methods rely on accurate knowledge of the blade-relative inflow velocity. As this is not known analytically because of turbine induction (which cannot be predicted accurately \textit{a priori}, at present), the estimation of virtual camber has a degree of uncertainty, particularly downstream. Acknowledging this fact,  we rely on the method proposed by Migliore and Wolfe \cite{Migliore1980} to assess magnitudes and trends of virtual geometry, as deviations due to induction surpass those due to the assumptions inherent to each alternative model. We refer the reader to Migliore and Wolfe \cite{Migliore1980} for a more detailed description of the transformation. As induction has a greater influence on the downstream portion of the blade rotation $(180<\theta<360)$, we expect the greatest deviations from the calculated virtual camber and $\alpha$ profiles to occur here. At high $\lambda$, the relative velocity of the blade rotation converges to the tangential velocity. Downstream induction accelerates this convergence through the reduction of the flow velocity within the turbine, which impinges on the downstream blade \cite{snortland2025downstream}. Consequently, the downstream portion of the rotation is expected to approach the virtual camber and incidence resembling those at $\lambda = \infty$ at lower rotation rates than would be needed for the upstream portion of the rotation. 

\begin{figure}
    \centering
    \includegraphics[width=0.99\textwidth]{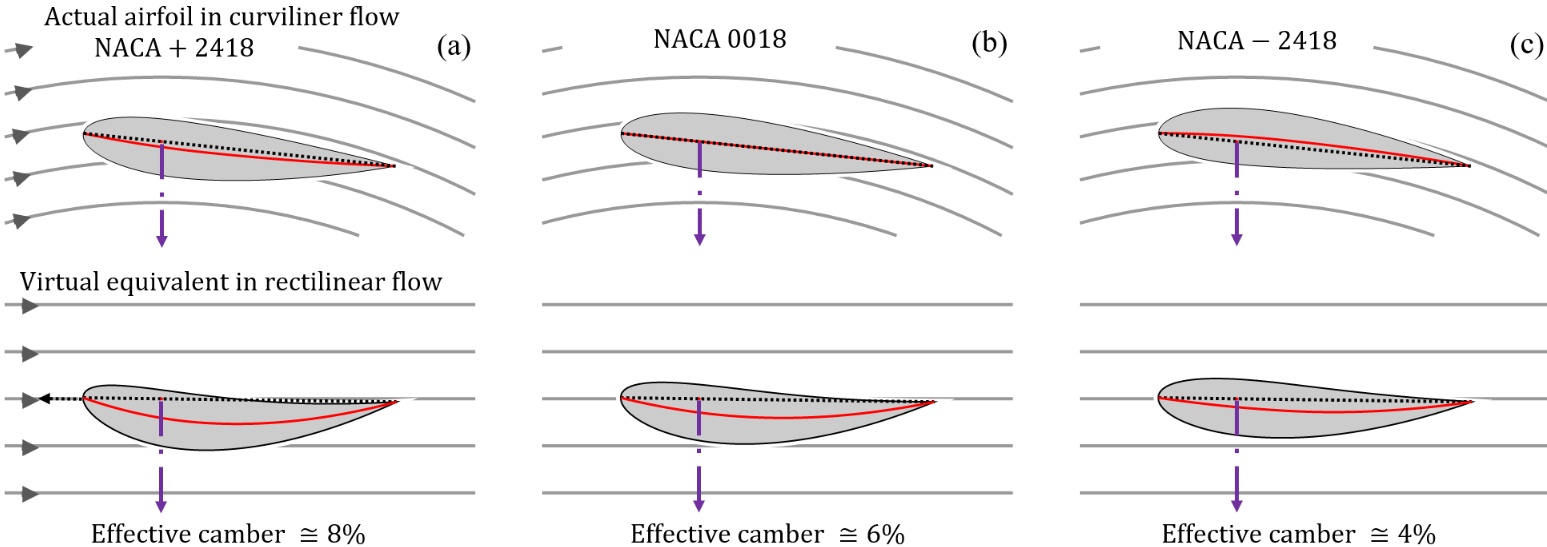}
    \caption{The turbine blades considered in this work transformed according to the method of Migliore and Wolfe \cite{Migliore1980}, from a curvilinear to equivalent rectilinear flow. Dotted black lines correspond to the chord line of the foils, (red) corresponds to the camber line, and purple represents the radial direction to the center of the turbine. The virtual incidence for all three foils was $\sim$ 7$^\circ$ which negates the preset pitch, resulting in a $\sim$1$^\circ$ inward pitch shown in the rectilinear equivalents.}
    \label{fig:camberstransformed}
\end{figure}

While virtual camber transformations have primarily been applied to symmetrical blades in the literature, the mathematical framework for extending these transformations to geometrically cambered blades is well-defined. Since the method of Migliore and Wolfe \cite{Migliore1980} assumes a thin airfoil for the transformation, the camber line is the primary input. The resulting virtual camber is primarily independent of the input geometric camber, and so the transformation can be thought of as a superposition of the foil's virtual and geometric camber (particularly for small cambers <$\pm$4\%). 
For foils with geometric camber centered between 0.4 and 0.6 $x/c$, the virtual camber line is smooth, and the effective camber magnitude is approximately the sum of the geometric camber and the virtual camber of a symmetrical foil.  This is illustrated in figure \ref{fig:camberstransformed} for the blade profiles studied in this work at the limit of $\lambda = \infty$. The virtual camber of the symmetrical foil (figure \ref{fig:camberstransformed}b) is seen to be 6\% while the positive and negative 2\% foils are shown to increase and decrease the total effective camber to 8\% and 4\% respectively.  For foils with more extreme positions of maximum camber, $x/c<0.4$ or $x/c>0.6$, the total camber can exhibit multiple maxima or a reflexive profile depending on the sign of the geometric camber, increasing the final profile complexity. We intentionally selected profiles with a position of max camber at $x/c = 0.4$ to avoid exploring such foils in the current study for this reason. 

\section{Results}
\label{Sec:Results}
\subsection{Exploration of Virtual Camber and Incidence Dependencies}
\begin{figure}
\centering
    \includegraphics[width=0.65\textwidth]{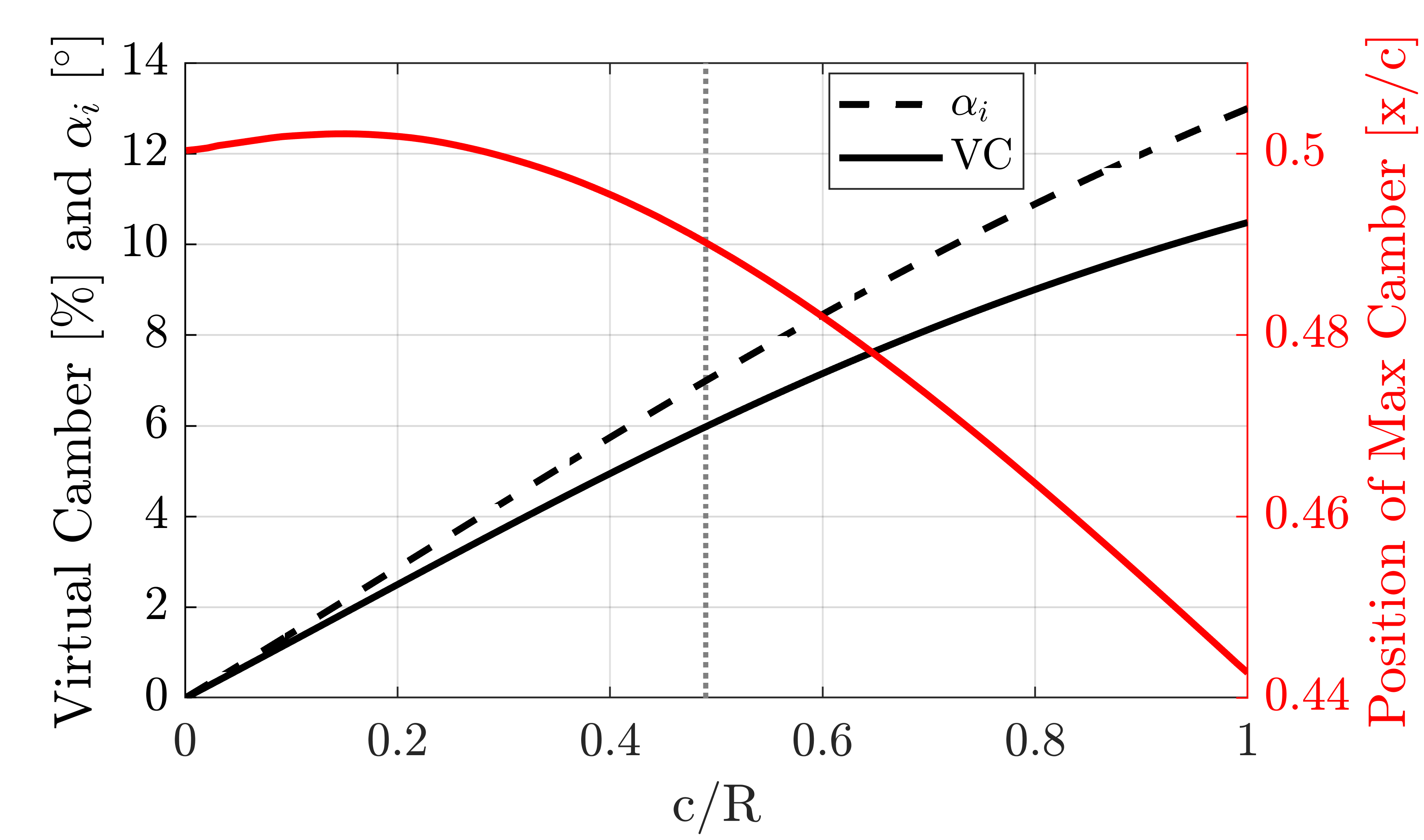}
    \caption{Impact of $c/R$ on virtual camber and $\alpha_i$, as well as the location of maximum virtual camber for $\lambda = \infty$. The vertical dotted line corresponds to the blade geometry explored in this work, $c/R = 0.49$.}
    \label{fig:VC_cRtrend}
\end{figure}%

We explore the variation of estimated curvature-induced virtual camber and incidence so that they can be compared to the geometric camber of the chosen airfoils. Throughout this analysis, we assume the blade mounting point to be at the quarter chord.  Figure \ref{fig:VC_cRtrend} shows the effects of $c/R$ on the virtual equivalent airfoil geometry in a purely rotational flow (i.e., at infinite $\lambda$, no inflow). The virtual incidence and camber are seen to vary in an approximately linear fashion with $c/R$, up to a maximum of 13$^\circ$ and 10\%, respectively. The position of maximum virtual camber is non-linear and varies between 0.44c and 0.50c. For $c/R = 0.49$, used by the turbines in the current study, a virtual camber of 6\% is observed, three times the maximum geometric camber that was tested. Similarly, the virtual incidence ($\alpha_i$) is 7$^\circ$ and opposes the preset pitch of $-6^\circ$. The position of the maximum virtual camber is $\sim$0.49 for the current study, close to the location of the maximum camber of the NACA-2418 foils. This geometry was explicitly chosen to be close to matching the virtual profile while also being a foil whose aerodynamic performance has been well documented. 

\begin{figure}
    \centering
    \includegraphics[width=0.85\textwidth]{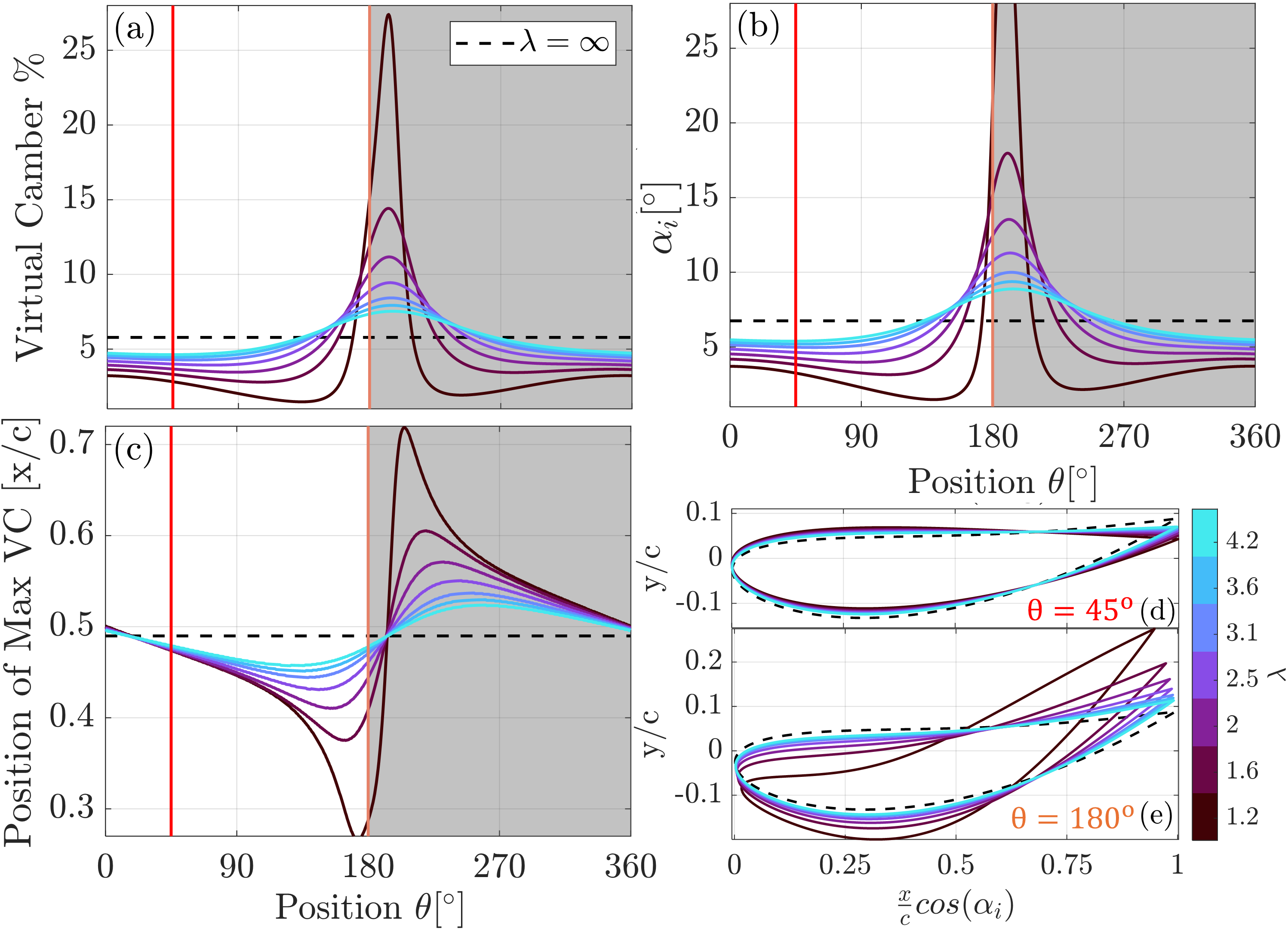}
    \caption{Trends of virtual camber (a), virtual incidence (b), and the position of max camber (c) for a range of $\lambda$, as a function of blade position, $\theta$. Calculated using the method of Migliore and Wolfe \cite{Migliore1980} for $c/R = 0.49$ and preset pitch of  $-$6$^\circ$ measured from a quarter chord mounting point. Cases with a purely rotational flow $(\lambda = \infty)$ are shown with the dashed black lines. The downstream region is shown by a gray background, where higher uncertainty exists and true behavior is expected to more closely match the $\lambda=\infty$ case. The associated virtual equivalent profiles for a NACA 0018 foil are shown for azimuthal positions of 45$^\circ$ (d) and 180$^\circ$ (e).}
    \label{fig:VC_TSRtrends}
\end{figure}

For a finite and constant $\lambda$, the virtual geometry varies as a function of blade position $\theta$, as explored in figure \ref{fig:VC_TSRtrends} for a $c/R$ of 0.49. While the virtual camber is consistently positive through the cycle, the greatest variations are seen to occur at low $\lambda$, where the inflow to the turbine has a greater influence. As such, expectations from steady-state aerodynamics suggest virtual camber would enhance upstream lift compared to a symmetrical blade at the same $\alpha$. 
The virtual incidence profiles (figure \ref{fig:VC_TSRtrends}b) follow those of virtual camber very closely, maintaining a positive value through the cycle, unlike the nominal $\alpha_n$ observed in \ref{fig:Coordinate}b which changes sign. This positive incidence corresponds to a nose-in pitch of the blade, increasing the upstream angle of attack and lift. For $\lambda>2$, the virtual camber and incidence increase throughout the cycle until just after $\theta = 180^\circ$ before diminishing, while lower $\lambda$ show a brief and moderate reduction of incidence and virtual camber upstream. The position of the maximum virtual camber, figure \ref{fig:VC_TSRtrends}c, moves from the mid-chord towards the blade's leading edge before dramatically shifting to the aft half of the foil for $\theta>180^\circ$. For the majority of the cycle, the resulting virtual airfoil (see examples in figures \ref{fig:VC_TSRtrends}(d,e)) appears similar to traditional NACA airfoils except under low $\lambda$ conditions and for phases near $\theta = 180^\circ$ where we see more extreme shapes due to a high rate of change in $\alpha_n(\theta)$. With increasing $\lambda$, virtual incidence and virtual camber increase upstream and asymptotically approach the steady-state $\lambda = \infty$ case (represented by the dashed horizontal lines in figure \ref{fig:VC_TSRtrends}). As induction in the downstream is neglected in the transformation method \cite{Migliore1980}, the virtual geometry predictions have higher uncertainty for $\theta>180^\circ$ and are expected to deviate toward the infinite result $\lambda$, due to lower inflow velocities at the blade, as noted previously.

\subsection{Influence of Geometric Camber on Turbine Performance}
\label{sec:ResultsPerf}

\begin{figure}
    \centering
    \includegraphics[width=1\textwidth]{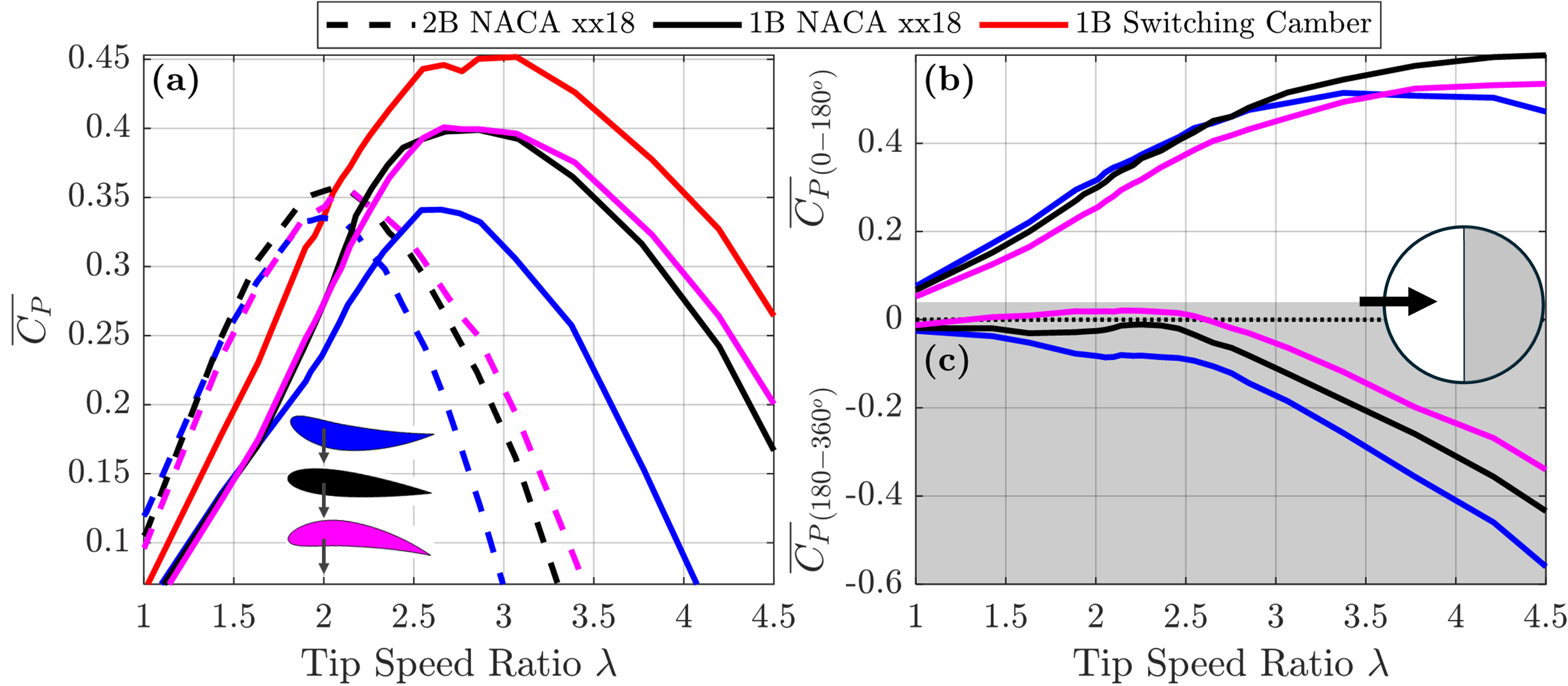}
    \caption{(a) Time-averaged performance of a one- (solid) and two-bladed (dashed) turbine as a function of rotation rate. (b) Time averaged upstream and (c) downstream performance of a single blade such that the sum is equal to performance of the whole turbine in (a). Positively cambered foils are shown in blue, symmetrical in black, and negative in pink. Arrows attached to the blade in the legend of (a) point toward the center of rotation. Red corresponds to the average performance of a hypothetical one-bladed turbine with the best performing profiles in the upstream portions of the rotation.}
    \label{fig:GCTimeAvg}
\end{figure}

The time-averaged performance of one- and two-bladed configurations are shown in figure \ref{fig:GCTimeAvg}a for the different blade cambers tested (symmetric and $\pm$2\% camber). It is revealed that the optimal geometric camber is dependent on the operating condition of the turbine for both configurations. For one- or two-blades, the symmetric or negatively cambered foils produce very similar time-averaged power output for most rotation rates, whereas the performance of the positively cambered foil produces less power at all but the lowest $\lambda$, contradicting earlier expectations that the positively cambered foil would give the best overall performance due to expected enhancements to upstream lift. Overall,  the negatively cambered foil performs best at the highest $\lambda$, and symmetric foils provide the best time-averaged performance at more moderate-to-low $\lambda$. 

Peak performance for one- and two-bladed turbines with symmetrical foils occurs at $\lambda$ values of 2.8 and 2, respectively and we see minimal variation due to geometric camber. For two-bladed turbines, the optimal $\lambda$ slightly decreases with increased camber, showing values of 2.15, 2.05, and 1.98 for negative, symmetrical, and positive cambered blades. While the one-bladed turbine shows similar trends overall, the difference in optimal tip-speed of symmetric and negatively cambered blades lies within experimental accuracy. 

The one-bladed data is used to isolate the upstream and downstream contributions to overall turbine performance for each camber (see figure \ref{fig:GCTimeAvg}b,c). Following the analysis of Snortland et al. \cite{snortland2025downstream}, these results further highlight the linkages between the downstream portion of the rotation and the optimal $\lambda$, with the optimal rotation rate governed by degradation in downstream performance observed at higher rotation rates. The upstream performance continues to increase past the optimum, and the downstream is approximately neutral (or slightly negative) until a critical $\lambda$ at which the performance decreases quickly and in an approximately linear fashion.

While the positively cambered foil has better performance than the negatively cambered foil in the upstream portion of rotation for lower $\lambda$, as expected, this is more than offset by diminished performance in the downstream sweep in almost all cases.  Negative camber is seen to reduce upstream performance, but the performance is much improved in the downstream region and even generates a small mount amount of positive power near optimal conditions. In other words, performance benefits in one region are generally offset to some extent within the other, with improvements in flow recovery in the downstream portion of the rotation appearing to be a key factor for this turbine. The upstream performance is also seen to peak at a much lower $\lambda$ for the positively cambered foil, significantly contributing to its poor overall performance.   
To provide a bound for potential improvements in turbine performance due to camber, we consider a hypothetical blade that changes camber through the cycle to match the best performing profile. The resulting time-averaged performance would be as shown in by the red profile in \ref{fig:GCTimeAvg}a. Such a blade would have between 10\% and 36\% improvement in efficiency compared to the symmetrical blade, with a higher potential for improvement at low $\lambda$ by percentage. Preliminary work has already designed airfoils with variable camber \citep{MorphingBlade2020}, which could be tested in cross-flow turbine applications to take full advantage of upstream/downstream preferences.

\begin{figure}
    \centering
    \includegraphics[width=1\textwidth]{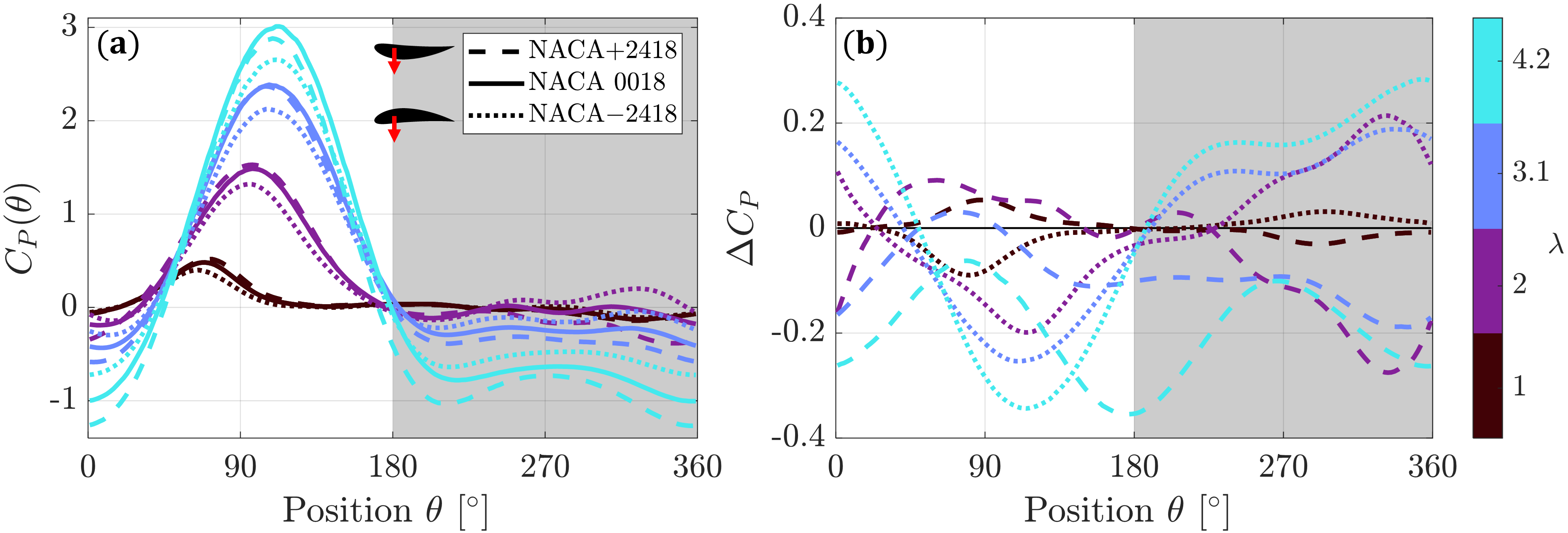}
    \caption{(a) Phase-averaged performance of the single-bladed turbine and (b) the difference in phase-averaged performance between cambered cases and the symmetrical-bladed turbine. Arrows attached to the blade in the legend of (a) point toward the center of rotation. 
    The colors correspond to $\lambda$ and line styles to blade geometry.}
    \label{fig:GCPhaseAvgPerf}
\end{figure}%

Further exploration of phase-averaged performance of the one-bladed turbine (see figure \ref{fig:GCPhaseAvgPerf}a) illustrates a more detailed comparison of performance throughout the cycle as a function of $\lambda$. By subtracting the symmetrical-bladed performance at equivalent $\lambda$'s the phase-dependent changes in performance due to different cambered geometries are highlighted in figure \ref{fig:GCPhaseAvgPerf}b.
As might be expected, it is seen that more positive geometrically cambered blades have a higher and slightly later upstream performance peak for $\lambda = 1-3$ (indicating stall delay), however for $\lambda > 3.4$ it is seen to fall off and become less then that of the symmetrically bladed turbin. The azimuthal location of the peak performance is shifted later in the rotation with increasing camber for below-optimal $\lambda$. In contrast, the phase of stall appears to be approximately independent of geometric camber above the optimal tips-speed. We see that previously observed decreases in the integrated upstream performance $(0<\theta<180)$ at high $\lambda$ for positive camber (seen in Fig \ref{fig:GCTimeAvg}b) is likely due a continuation of poor flow recovery as the blade passes  $\theta =0^\circ$, delaying the start of the lift-generating process to later in the cycle. In the downstream, the performance becomes progressively more negative at the highest rotation rates. This effect is amplified for the more positively cambered foils. In contrast, the negative foil shows a positive secondary downstream peak around $\theta=315^\circ$ at moderate $\lambda$ (i.e., $\lambda = 2$), suggesting a secondary stall event that produces lift. Furthermore, as the geometric camber increases, more detrimental pitching moment values are expected, which lead to performance degradation at high TSR. This observation builds on the findings of Snortland et al.\cite{SNORTLAND2025Force} and provides further insight into how camber and TSR impact performance.

\subsubsection{Influence of Geometric Camber on Turbine Loading}

\begin{figure}
    \includegraphics[width=0.95\textwidth]{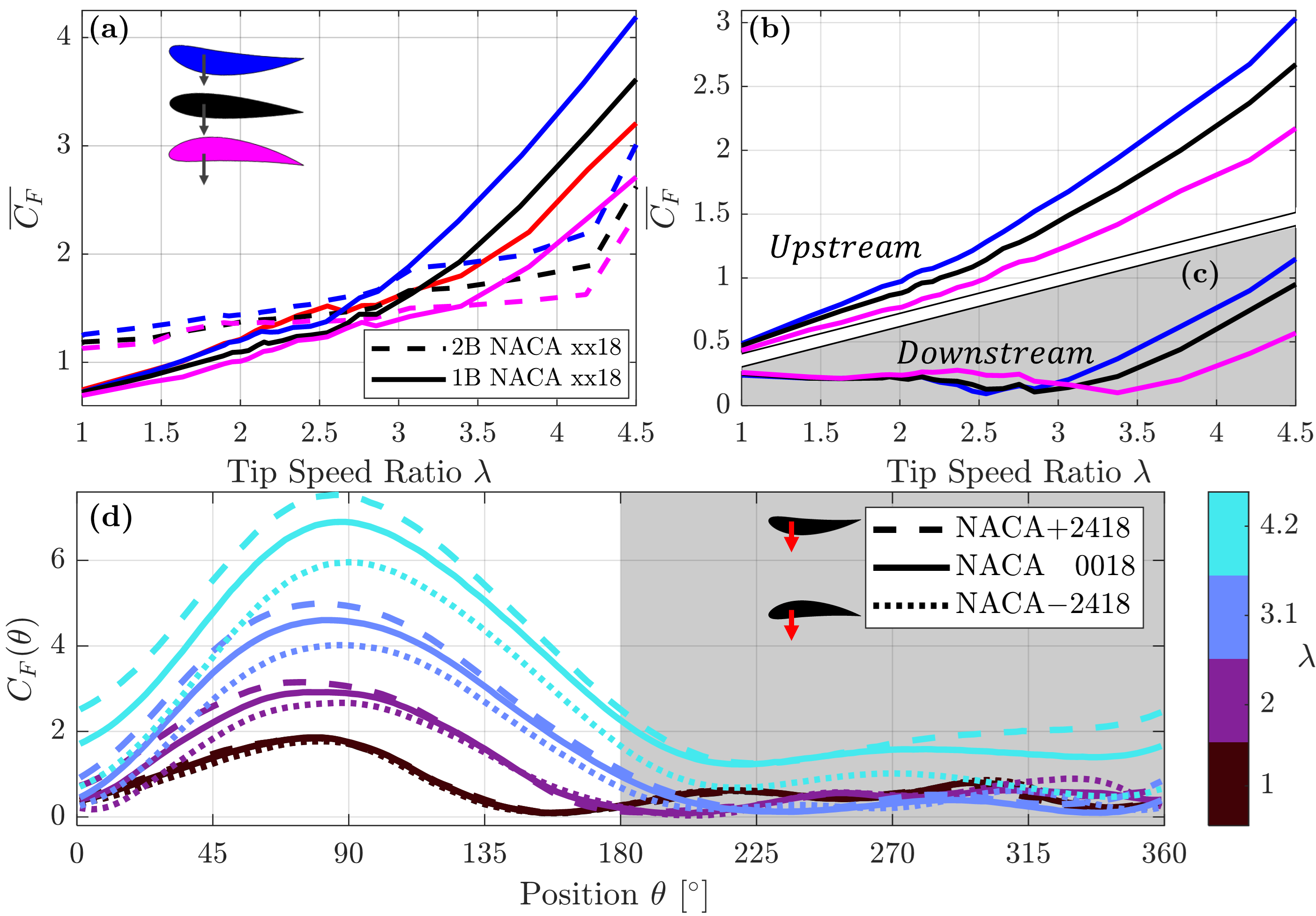}
    \caption{(a) Time-averaged one- and two-bladed in-plane loading. (b) Upstream and (c) downstream average force on a one-bladed turbine. (d) Single-blade, phase-averaged horizontal force, with colors corresponding to $\lambda$ and line style representing blade geometry. Arrows attached to the exaggerated blade in the legend of (a) and (d) point toward the center of rotation.}
    \label{fig:GCTime_PhaseForce}
\end{figure}

The time-averaged in-plane, support-subtracted loading on one- and two-bladed configurations are shown in figure \ref{fig:GCTime_PhaseForce}a.  
While the optimal geometric camber for performance is shown to be dependent on the rotation rate, loading is much less nuanced. The mean and peak turbine loading is seen to be larger for increasingly positive camber across all tip-speed ratios, and the difference in average loading between the tested blade geometries grows with rotation rate. Conversely, turbine loading is also seen to be reduced with a negative geometric camber. 

Figure \ref{fig:GCTime_PhaseForce}(b-c), reveals that the loading in the upstream half of the rotation increases directly with increasingly more positive camber, while the camber that produces the greatest average downstream loading shows a smaller $\lambda$ dependence.  For $\lambda<2$, downstream loading is seen to be relatively independent of camber profile. Near the $\lambda$ of optimal performance ($\lambda = 2.5-3$ for the one-bladed turbine), the negative camber sees the highest mean loading. 
And as $\lambda$ increases above a value of three, the average downstream loading starts to increase, with higher camber seeing consistently higher mean loads once again. This is expected to occur due to the higher drag and radial lift generation that is generated for the positive camber as the blades approach steady-state conditions. These results suggest that, at the optimal one and two-bladed $\lambda$, the average in-plane blade loading can be reduced by (6.7\% and 0.8\%) respectively by changing from a symmetrical blade to the negatively cambered one. 

We can similarly examine the maximum in-plane blade loading by examining the phase-averaged variation of the one-bladed turbine in figure \ref{fig:GCTime_PhaseForce}d. We see that more positively cambered blade increases the maximum loading (which occurs in the upstream portion of the rotation). Conversely, at the highest $\lambda$ tested ($\lambda$=4.5), the negatively cambered blade produced a peak blade level loading 21\% less than the positively cambered foil and 13\% less than the symmetrical. At the optimum $\lambda$, the maximum loading on the negatively cambered blades was 19\% less than the positively cambered turbine and 11\% less than the symmetrical. In addition, the position of maximum force shifts later in the cycle for lower geometric camber, opposing what is observed for $C_P$, where a higher camber shows the most delayed peak. 

Section \ref{sec:ResultsPerf} discussed the performance of a hypothetical blade that can switch camber direction to obtain beneficial characteristics of both the upstream of downstream regions.  The influence of such a variable foil on average blade loading is modeled in figure \ref{fig:GCTime_PhaseForce}a (redline), showing up to a 19\% increase at or below the optimal $\lambda$ but a 7-11\% reduction beyond peak $\lambda$ compared to a traditional symmetrical foil, with the largest reduction at high $\lambda$. Overall, the loading on such an actively cambering foil is only slightly higher than a fixed negative camber foil while further enhancing performance.

\subsection{Flow Field Analysis for $\lambda$ = 2}

\label{Sec:PIV}

\begin{figure}
    \centering
    \includegraphics[width=0.99\textwidth]{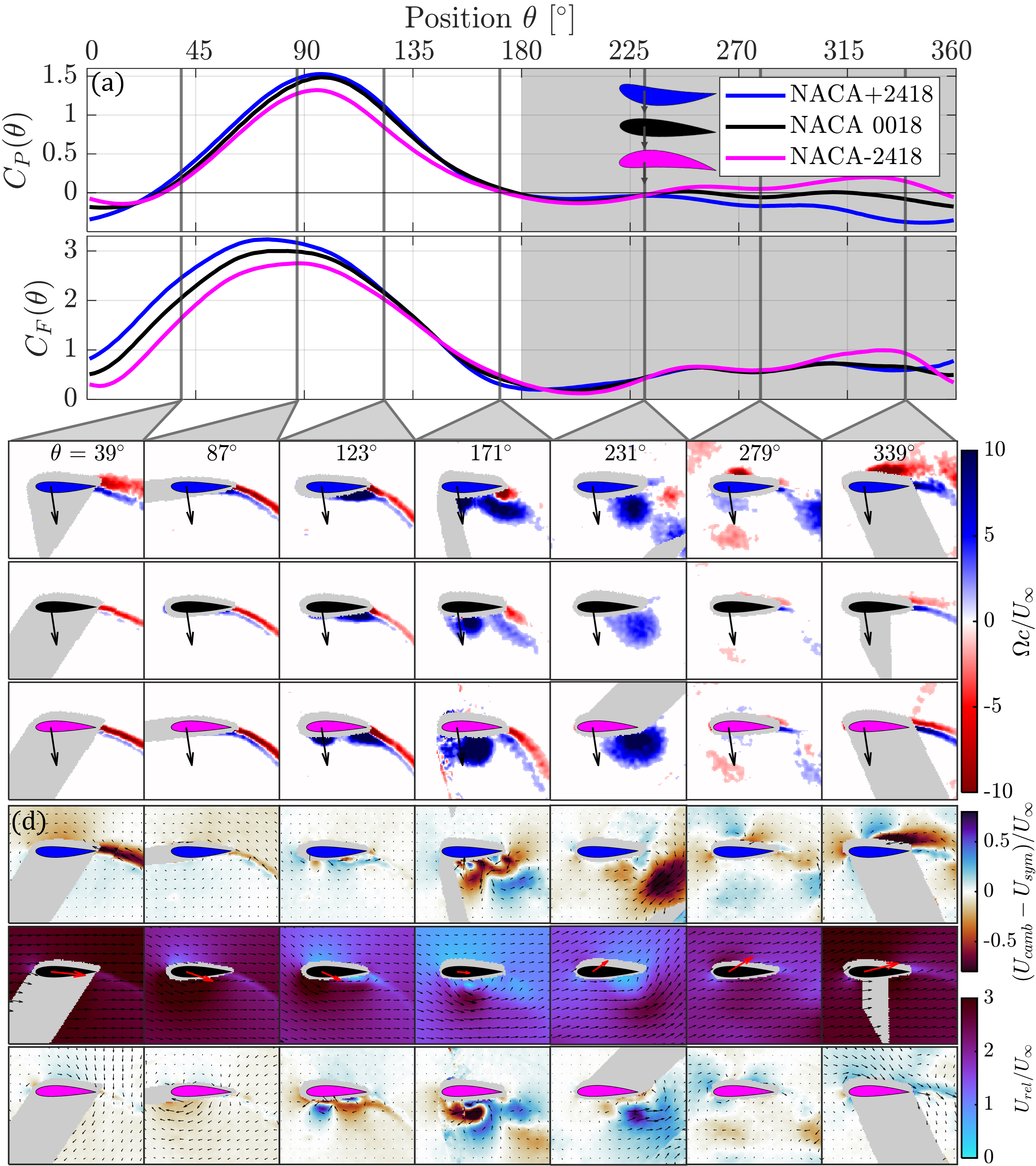}
    \caption{Comparison of one-bladed phase-averaged performance and near-blade hydrodynamic structures across blade profiles at $\lambda = 2$. Support subtracted phase-averaged (a)  $C_P$, (b) $C_F$. (c) Near blade vorticity fields with black arrow indicating radial direction, and (d) shows both blade relative velocity fields and difference in relative velocity between the symmetrical case and cambered at matching conditions. In (d), the red arrows attached to the blades indicate the nominal blade-relative velocity independent of geometric camber. Arrows attached to the blades in the legend of (a) point toward the center of rotation. }
    \label{fig:CambPIV}
\end{figure}

Exploring the phase-averaged flow fields allows direct linkages to be made with the performance and loading trends observed above. The interconnections between shed vorticity and the phases and types of dynamic stall and reattachment are explored. In figure \ref{fig:CambPIV}, we compare phase-averaged $C_P$, the in-plane force coefficient, $C_{F}$, and phase-averaged near-blade vorticity and blade-relative velocity fields for each airfoil at $\lambda = 2$, which is close to optimal for the two-bladed turbine. The PIV results are presented in a blade-centric coordinate system. Single-bladed results are shown here to allow direct connection between blade-level performance and observed hydrodynamics. The time-averaged performance of the symmetric and negatively cambered one-bladed turbine at $\lambda =2$ is almost equivalent, while the positively cambered performs significantly worse (figure \ref{fig:GCTimeAvg}a). The two-bladed data shows similar trends in both efficiency and force at this rotation rate, so we expect the one-bladed hydrodynamic results to also be representative of the two-bladed configuration at these conditions.  

The vorticity fields for the symmetric blade (black foils in figure \ref{fig:CambPIV}(c)) are typical of documented trends of flow separation throughout the cycle \cite{le2022dynamic}. Early in the cycle $\theta = 0-90^\circ$, attached flow is observed along the blade, where angle of attack increases and thus power production and forces. As the blade progresses to $\theta\approx120^\circ$, visible dynamic stall features are observed,  with growing leading and trailing edge vortex development (blue) near the blade, which induces an unfavorable change do the direction of lift relative to torque generation resulting in a loss of power production. The dynamic stall vortex  (blue)  originating at the blade tip grows through the cycle, and advects along the blade before shedding off, $\theta = [120- 225^\circ]$, and is seen to cause a dip in power production. As the blade progresses, the blade relative velocity is seen to increase and flow reattachment occurs as the uniform shear wake is observed near $\theta = 270^\circ$ to start the process over.

In contrast, the positively cambered blade sees both a different stall and recovery process compared to the symmetric blade. Early in the cycle at the vorticity along the blade appears almost identical to the symmetric blade at $\theta = 90^\circ$: however, differences in $C_P$ are still observed; likely linked to the ability of camber to enhance lift. This difference in lift can be inferred by the velocities being lower on the outer surface of the foil and accelerated along the inner in figure \ref{fig:CambPIV}d. The positively cambered foil is seen to have a slightly delayed stall compared to the symmetric foil, and it starts more prominently at the trailing edge. This delay in stall, along with accelerated relative velocity on the inside of the blade, allows for higher power and force generation upstream, matching the performance measurements $\theta =40-120^\circ$. The positively cambered blade also exhibits different dynamic stall behavior, with both significant leading and trailing edge vorticity observed ($\theta=170^\circ$). A larger stall vortex is seen to form near the trailing edge, along with a very small counter-clockwise vortex at the blade trailing edge (red). This small vortex is less noticeable in the presence of the stronger vortices of the opposite sign close downstream, resulting in a less severe shedding of what appears to be two clockwise (blue) vortices rather than the single shed vortex observed for the symmetric foil. The later shedding of vorticity by the more positively cambered foil leads to increased lift later in the cycle, as resulting blade-relative velocities are lower on the underside of the blade for $\theta=170-220^\circ$. The stronger trailing edge vortex formation during dynamic stall due to positive camber follows the limited observations in prior work with pitching foils \cite{mccroskey1981phenomenon,Ouro2018,Shum2024}, but this is the first time this has been observed experimentally while holding other airfoil design parameters constant in the rotating cross-flow turbine frame. 
In addition, in the the downstream portion of the rotation the positively cambered blade shows poor reattachment and sheds a secondary red vortex on the outside of the blade between $\theta = 280-340$ that was not present in the symmetrical case. The poor downstream performance of the positive camber prevents flow reattachment till $\theta=40^\circ$, leading to a broader wake and power degradation between $\theta = 270-40^\circ$. 

The negatively cambered blade sees an opposite trend in both recovery speed and stall strength, relative to the positively cambered blade. While at $90^\circ$ the vorticity along the blade is broadly similar to the other blades, the relative velocity on the inside of the blade is diminished in comparison to the symmetrical cases, resulting in the reduction of power production and loading. During the power stroke, the negatively cambered blade shows a notably larger clockwise leading edge stall vortex ($\theta=120^\circ$) compared to the symmetrical blade, matching observations by McCroskey \cite{mccroskey1981phenomenon} where less cambered foils (effective camber in this application) tend to exhibit more of a leading edge stall. This more dramatic stall leads to a loss of lift and force along the blade between $180 \text{ and } 225^\circ$ and is reflected in the force and performance data here. However, once this vortex is convected out of frame, the flow is largely attached on both sides of the foil leading to lift generation and positive performance ($300^\circ<\theta<360^\circ$) and a coupled rise in forces relative to the more positively cambered blades in the latter portions of the cycle. 

Across the three blade profiles we observe that there are non-linear changes to stall and recovery processes. While the strength and phases or vortex separation and reattachment are altered, greater changes are also present. Lower cambered blades promote stall closer to the leading edge, while the opposite is true for highly cambered blades. Secondary downstream stall events are also observed for the positive camber. Although the observed changes to stall dynamics with camber have seen some exploration in pitching foils, there remains no general model for the effect of camber on dynamic stall. Moreover, in the rotating frame of these turbines the influence of increasing virtual and geometric camber may influence stall differently. As a result the use of effective camber (the sum of virtual and geometric camber) as a design parameter may be overly simplistic, despite its utility.

\subsection{Exploring a wider range of geometric camber through simulation}
   
In an effort to estimate the optimal camber for this foil, we now explore the sensitivity of performance and loading to a wider range of geometric camber ($-$5\% to $+$3\%) at a constant $\lambda = 2$ through validated URANS simulations of a one-bladed configuration.

\begin{figure}
    \centering
    \includegraphics[width=.65\textwidth]{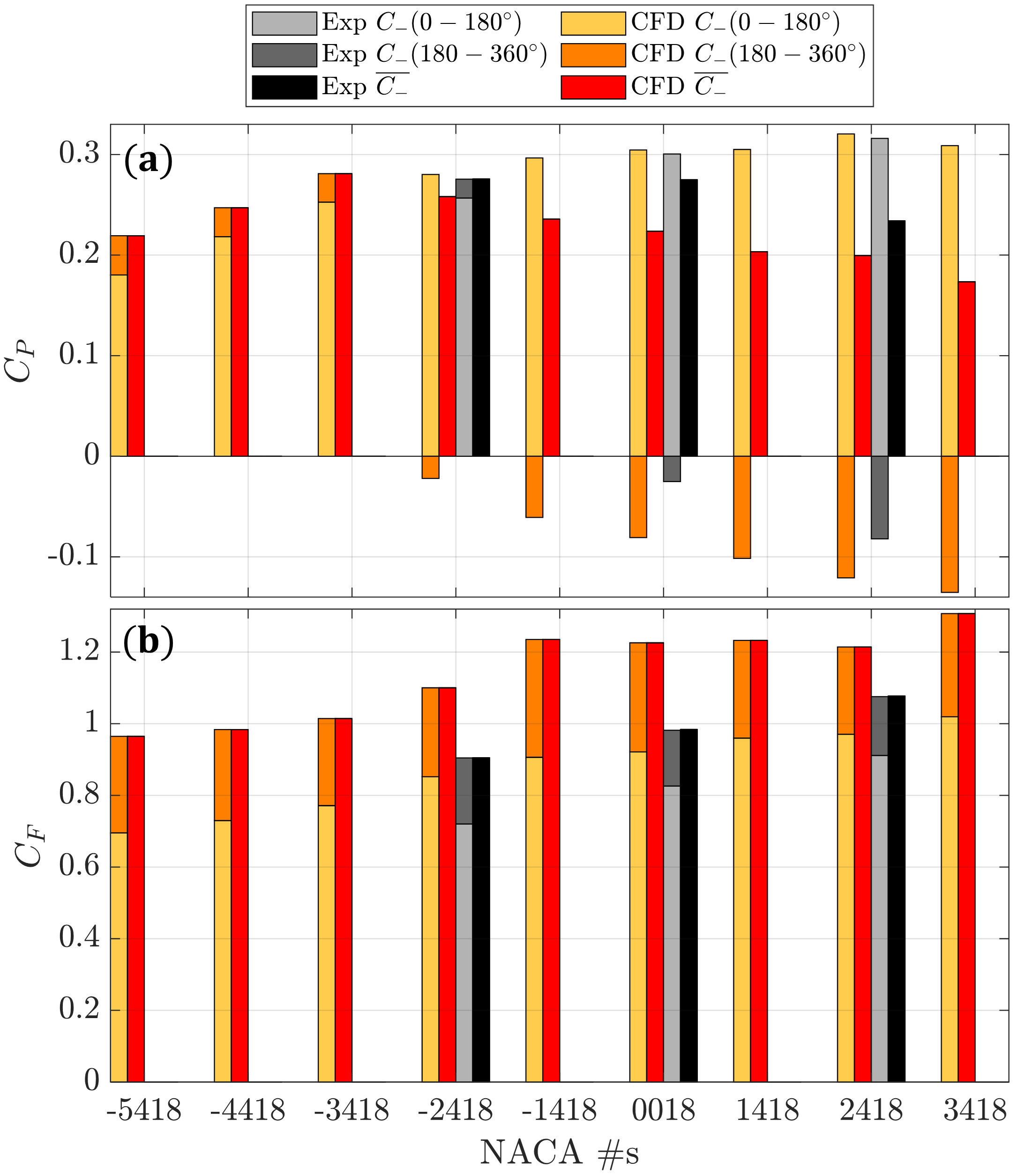}
  \caption{(a) Time-averaged $\overline{C_P}$ and, (b) $\overline{C_F}$ and their contributions from the upstream and downstream portions of the rotation over an expanded range of blade camber.}
    \label{fig:CFD_barchart}
\end{figure}

Computational time-averaged performance and loading, as well as the relative contributions from the upstream and downstream portions of the rotation, are shown in figure \ref{fig:CFD_barchart}. While the proportion of the performance derived from the upstream and downstream parts of the rotation varies slightly between the CFD and experimental results, both show similar trade-offs with camber, and computational trends broadly align with experimental findings. Negative geometric camber is seen to increase downstream performance across all tested foils, and positive camber shows improved upstream performance for camber of less than $+$2\%, at which point the upstream performance peaks and begins to reduce. Once again, improvements to upstream performance are observed to come at the cost of downstream performance. The loading trends are also similar to experiments despite overall forces being slightly higher in the simulations, potentially due to their two-dimensional nature. Overall, the in-plane $\overline{C_F}$ shows a consistent increase with increasing camber, primarily from the upstream contribution, with a slight, unexplained dip at $+$ 2\% before continuing to increase. The rate at which loading improves with negative camber is seen to reduce at approximately $-$4\% camber, but is still decreasing at the minimum camber tested of $-$5\%. 

The simulations suggest that the camber of optimal performance for a one-bladed turbine with $c/R= 0.49$, at $\lambda = 2$, is approximately $-$3\%, and that the loading is also significantly reduced for this profile. This amount of geometric camber corresponds to $\sim50\%$ of the virtual camber, leading to a total camber (virtual plus geometric) of $+3\%$, assuming the limiting case of $\lambda=\infty$ is representative of the virtual camber effect. Comparing the loading and power generation of this case to the simulated symmetric NACA 0018 performance shows a 25.6\% increase in $\overline{C_P}$ and 17.2\% decrease in $\overline{C_F}$ at the same $\lambda$.

\section{Discussion and Conclusions}

This work has shown that ideal blade camber depends on $\lambda$, with a higher camber favored (if only slightly) at low rotation rates, where upstream performance is critical to overall power production, and a lower camber is favored at higher rotation rates, where the downstream region and blade stall recovery dominate the overall performance. This is in contrast to the assumption based on linear airfoil behavior that positive camber should enhance total performance due to improved lift in the power stroke, independent of rotation rate. Explorations of phase-averaged performance indicate that changing camber exhibits trade-offs in performance between the upstream and downstream regions. Increasing camber, which has been shown to benefit the upstream region, is coupled with a drop in performance at high tip-speed-ratios. 

Results from this study indicate performance can be increased at high tip-speed by reducing the magnitude of effective camber, which reduces both detrimental drag and pitching moment. This is consistent with recent work by Snortland et al.\cite{snortland2025downstream} showing that at
high tip-speed performance is limited by downstream drag and the pitching moment, as the blade-relative velocity tends toward the tangential velocity, and the dynamics become more steady-state. Similarly, loading is dominated by lift production normal to the blade in this regime. The speed of downstream flow recovery, seen to be vastly different between different cambered blades, would further enhance the disparity. The hypothesis that enhancements to blade lift in the upstream portions of the rotation would enhance overall performance turns out only to be valid when the downstream flow recovery has a reduced influence (and only to a limited extent), such as seen at lower tip-speeds. Overall performance is a subtle balance between upstream power production and downstream flow recovery, and the optimum camber will depend on final turbine operational conditions. In summary, our results reveal that, while optimal camber is tip-speed dependent, an effective camber of approximately $+3$\% (or $-$3\% geometric camber) may be optimal for power generation at this $c/R$ at a tip-speed ratio close to optimal for the two-bladed turbine ($\lambda=2$). This corresponds to a geometric camber around 50\% of the virtual component. Any global parameters that alter the phase of upstream blade stall or flow induction (such as blade pitch, confinement or turbine solidity, for example) might be expected to change this balance and therefore the optimum geometric camber for a given tip-speed ratio. 

This work highlights the significant potential of negative cambered blades to enhance turbine efficiency and loading, with power output improvements of up to 20\% beyond the optimal tip-speed of the two-bladed turbine while reducing peak and average loading by up to 9\% and 14\%, respectively. If the blade could dynamically switch between positive and negative camber through the cycle, the performance enhancement is estimated to be even greater at 11-71\% of that of the symmetric foil (depending on tip-speed), while reducing loading by 7-11\%. It should be noted that these percentages are for the chosen $c/R$ of 0.49, and further explorations will be needed to determine if these trends hold for a broader range of relative blade curvature. 

 More negatively cambered blades show broad benefits that may prove crucial to practical turbine deployments. Reduced maximum loading and torque—achieved through a weaker power stroke combined with faster flow recovery during the downstream rotation—will produce smoother power delivery and structural advantages. Since loading and structural design considerations are key in reducing the fixed costs of turbine system designs, these insights are critical for optimizing efficiency and cost-effectiveness. In addition, the performance curves of the negatively cambered blades showed a expanded range of $\lambda$'s which produced positive power, increasing the area under the $C_P$ - $\lambda$ curve. This expands the potential for over-speed control implementation during region three control, where loading needs to be shed. This broadening of the operation domain is also critical in applications where large-scale turbulence is present and inflow velocities fluctuate significantly, suggesting that negatively cambered blades may be highly valuable in many field deployments. Furthermore, as more negatively cambered blades are utilized, power generation shifts from the upstream peak to the downstream region, reducing the peak-to-average ratio and the variability of power generation, which has beneficial implications for generator efficiency and grid connection.

While tip-speed dependent, the results of the current study suggest that a positive total camber (virtual plus geometric) is likely desired as, at the $c/R = 0.49$ explored in this work, the magnitude of virtual camber ($+6\%$), is greater than the $-$2\% to $-$3\% geometric camber that was found to be optimal at a tip-speed ratio of two. Previous studies have found negative or positive geometric camber to be optimal. While variations in tip-speed or analysis approach may partially explain this, we hypothesize that differences in the magnitude of the virtual camber contribute significantly to this reported variation. If differences in flow curvature ($c/R$) and resulting virtual camber, calculated for an infinite tip speed, are taken into account (and terminology and sign conventions modified to match the current paper), results from earlier cambered studies appear more consistent. The optimal foil of Kirke \cite{Kirke1991}, from a self-starting perspective, was a NACA 4418, resulting in an effective camber of $+$5.6\% ($c/R$=0.123, virtual camber = 1.6\%). The optimal profile from Bausas and Danao\cite{Danao2016MoreCambers} ($c/R$ = 0.1, virtual camber = 1.3\%) had an effective camber of $-$0.2\%. Zanforlin\cite{Zanforlin2023} found an optimal profile ($c/R = 0.33$, virtual camber = 2.1\%) had an effective camber of 4.1\%. Differences in the optimal value likely still stem from variations in the objective, such as power optimization\cite{Danao2016MoreCambers}, self-starting\cite{Kirke1991}, or wake recovery\cite{Zanforlin2023}, as well as the tip-speeds examined, or the overall blade profile details explored (such as thickness). Direct comparison with prior work, however, remains challenging due to differences in preset pitch, rotation rates, and validation or modeling approaches. Simplified models, such as DMST and CFD, often fail to accurately capture the complex, unsteady separation present in cross-flow turbines, especially given the limited availability of experimental data for validation. Despite this, the relatively small differences in optimal camber magnitude combined with our results suggest that an overall net positive camber may be beneficial across the design space. Furthermore, our work highlights the utility of a negative cambered blade to reduce excessive virtual camber and enhance efficiency.

The effective camber, summing the virtual and geometric components, thus appears to be a useful tool to compare between turbines at different $c/R$ and to reconcile past studies. However, we must be cautious. PIV results from the current study show that the shedding and reattachment processes change in a highly non-linear way with alternations in geometric camber (promoting either leading or trailing edge stall, for example).  It is thought that virtual camber is unlikely to have an identical influence as geometric camber on local pressure fields, flow separation, and reattachment processes. As a result, the effective camber, while a promising metric, should be used only for guidance at present until more conclusive results about its generality can be obtained. In addition, further generalization of this guidance for turbines of arbitrary $c/R$ will require consideration of the influences of blade virtual incidence and pre-set pitch on the resulting optimum, as these two variables are also likely to change significantly with $c/R$.

In conclusion,  blade camber is seen to have a significant impact on cross-flow turbine performance, loading, and hydrodynamics, providing a basis for tailoring cross-flow turbine blades to specific needs.
Collectively, our work highlights the complex interplay of blade camber and overall turbine geometry and operation, with the effective camber appearing to account for the combined influences of virtual and geometric camber, despite nonlinearities in flow field response to changes in geometric camber. It is demonstrated that the operating range of tip-speeds and loading should also be considered, with multifaceted design implications that extend beyond turbine efficiency, such as power quality and structural considerations.




\bmsection*{Conflict of interest}

The authors declare no potential conflict of interest.

\bmsection*{Data Availability}
The one-bladed experimental performance and PIV data from sections \ref{sec:ResultsPerf} and \ref{Sec:PIV} are available online via the Marine and Hydrokinetic Data Repository (MHKDR) at \url{https://mhkdr.openei.org/submissions/610}. Additionally, loading, computational, and two-bladed data are available from the corresponding author upon reasonable request.

\bibliography{cas-refs}
\appendix

\bmsection{Details of Support Subtraction}
\label{sec:Supplimental}
\vspace*{12pt}

Detailed description of support structure efficiency and loading losses.

To evaluate the phase-averaged blade-level forces of a one-bladed turbine, it is necessary to subtract the inertial loading, which arises from an offset in the axis of rotation and center of mass. This subtraction was conducted on the $F_X$ and $F_Y$ independently using the method proposed by Snortland et al. \cite{SNORTLAND2025Force}, using,
\begin{equation*}
\begin{aligned}
    F_X = F_{X,m} + mr_g(\omega^2sin(\theta-\delta\theta)) \\
    F_Y = F_{Y,m} - mr_g(\omega^2cos(\theta-\delta\theta)),
\end{aligned}
\end{equation*}
where $_m$ corresponds to the load cell measured quantities and $mr_g$ represents the inertial effect through the product of the rotating mass and radius to the center of mass, and $\delta\theta$ corresponds to the azimuthal offset of center of mass relative to the quarter chord. Supplemental experiments conducted in air to back out the $mr_g$ and $\delta\theta$ terms of our turbine. In these supplemental experiments, the velocity of the turbine was sinusoidally varied as a function of position. By assuming drag to be negligible in air, Newton's second law is applied and inertial constants are solved for as detailed in \cite{SNORTLAND2025Force}.
This inertial subtraction was not conducted for the axisymmetric experiments (i.e., for two-bladed turbines and support structure cases) as $mr_g$ is assumed to be negligible for a balanced configuration. 

After inertial effects were removed as necessary, support losses from the struts and central shaft were subtracted to isolate blade-level loading using the following approach. To isolate blade level $C_P$ and $C_{F_X}$,
\begin{equation*}
\begin{aligned}
                C_{[P,F_X],blade}(\theta) = C_{[P,F_X],turb}(\theta) - [C_{[P,F_X],struts}(\theta)-C_{[P,F_X],shaft}(\theta)]\\
                -C_{[P,F_X],shaft}(\theta)\frac{h-s}{h}
\end{aligned}
\end{equation*}%
was used, where subscripts $_{turb}$, $_{struts}$, and $_{shaft}$ correspond to full turbine experiments, strut and shaft supplemental data, and shaft only cases, respectively. This approach subtracts the effects of the struts only as well as the shaft influence outside of the turbine area ($\frac{h-s}{h}$) and assumes that due to induction, the torque and force in the streamwise direction on the shaft are negligible inside of the rotor. However, the lateral force was subtracted using

\begin{equation*}
                C_{F_Y,blade}(\theta) = C_{F_Y,turb}(\theta) - C_{F_Y,struts}(\theta)
\end{equation*}%
as it has been shown to have a better collapse \cite{SNORTLAND2025Force}.

To calculate the in-plane force magnitude, the component blade-level forces are then combined using, 

\begin{equation*}
                C_{Force,blade}(\theta) =\sqrt{C_{F_X,blade}^2(\theta) +C_{F_Y,blade}^2(\theta)}
\end{equation*}.%

 While the above equations are presented for phase-averaged performance as a function of $\theta$, the same equations were utilized for the calculation of time-averaged behavior by simply substituting functions of $\theta$ for their time-averaged equivalent (i.e. $C_{F_x,turb}(\theta) \rightarrow \overline{C_{F_x,turb}}$).

\end{document}